\documentclass[conf]{new-aiaa}
\usepackage[utf8]{inputenc}

\usepackage{graphicx}
\usepackage{amsmath}
\usepackage[version=4]{mhchem}
\usepackage{siunitx}
\usepackage{longtable,tabularx}
\usepackage{caption}    % For better captions

\usepackage{amssymb}

\usepackage{subfig}  % instead of subcaption
\usepackage{float}

\title{Adaptive Formation Control Strategy for Teams of Unmanned Vehicles under Complete Uncertainty}

\author{Maryam Norouzi \footnote{PhD Student, Department of Mechanical, Industrial and Systems Engineering, University of Rhode Island, USA}}
\affil{University of Rhode Island, Kingston, RI 02881, USA.}
\author{ Mingxi Zhou,\footnote{Associate Professor, Graduate School of Oceanography.}}
\affil{University of Rhode Island, Kingston, RI 02881, USA.}
\author{Chengzhi Yuan \footnote{Associate Professor, Graduate School of Oceanography, Industrial and Systems Engineering, University of Rhode Island, USA.} \footnote {Corresponding author}.}
\affil{University of Rhode Island, Kingston, RI 02881, USA.}

\begin{document}

\maketitle
\renewcommand*{\thefootnote}{\fnsymbol{footnote}} % optional: to use symbols instead of numbers
\footnotetext[0]{This work was supported by the National Science Foundation under Grant CMMI-2154901.}
\renewcommand*{\thefootnote}{\arabic{footnote}} % restore numbering style (if needed later)

\begin{abstract}
Modern unmanned systems, including aerial, terrestrial, and underwater vehicles, are increasingly utilized in dynamic and unpredictable environments, where the presence of modeling uncertainties necessitates the development of robust and adaptive control strategies. In this work, we address the formation control problem for a team of unmanned systems with completely uncertain dynamics under a virtual leader-following framework. We propose a novel cooperative adaptive formation control algorithm, designed using artificial neural networks to achieve accurate formation tracking. The effectiveness of the proposed control strategy is established through rigorous theoretical analysis, which guarantees uniform ultimate boundedness of the overall system and exponential convergence of the tracking errors to a small neighborhood of zero.  Numerical simulations further validate the effectiveness of the proposed formation control algorithm, demonstrating that the followers accurately track the desired formation trajectory relative to the leader, even in the presence of complete system uncertainties. This work suggests potential application in coordinating multiple unmanned airships for tasks such as persistent aerial surveillance, atmospheric data collection, and wide-area communication support, where adaptability to time-varying and uncertain dynamics is essential. 
\end{abstract}

% \section{Nomenclature}
% {\renewcommand\arraystretch{1.0}
% \noindent\begin{longtable*}{@{}l @{\quad=\quad} l@{}}
% $A$  & amplitude of oscillation \\
% $a$ &    cylinder diameter \\
% $C_p$& pressure coefficient \\
% $Cx$ & force coefficient in the \textit{x} direction \\
% $Cy$ & force coefficient in the \textit{y} direction \\
% c   & chord \\
% d$t$ & time step \\
% $Fx$ & $X$ component of the resultant pressure force acting on the vehicle \\
% $Fy$ & $Y$ component of the resultant pressure force acting on the vehicle \\
% $f, g$   & generic functions \\
% $h$  & height \\
% $i$  & time index during navigation \\
% $j$  & waypoint index \\
% $K$  & trailing-edge (TE) nondimensional angular deflection rate
% \end{longtable*}}

\section{Introduction}
Over the past decades, robotics has been applied in various fields, including industry \cite{c01, gholampour2025mass}, healthcare \cite{c02, mohammadi2024development}, assistive technology \cite{c03, salehimrl}, and more recently in aerial platforms, including unmanned airships, which are designed for low-speed, long-endurance missions in applications like surveillance, environmental monitoring, and communication support \cite{c027}. These advancements have collectively contributed to increased efficiency and improved quality of life. Within this broader domain, multi-agent systems (MASs) have emerged as a powerful approach for solving complex real-world problems, thanks to their efficiency, scalability, and adaptability. MASs refer to systems consisting of multiple agents that collectively perform complex tasks. The presence of multiple agents allows for the dividing of a complex task into smaller and manageable sub-tasks, where each agent handles a portion of the overall objective \cite{c04}. In many robotic applications, agents must adhere to a specific formation. Formation in multi-agent systems refers to maintaining a predefined geometrical configuration among agents—such as drones, robots, or autonomous vehicles—as they collaboratively execute coordinated tasks \cite{c014}. Achieving this requires precise coordination, collision avoidance, and efficient navigation around obstacles \cite{c05,c06,c07}. For instance, in forest-based search-and-rescue missions, drones must navigate through cluttered terrain while operating in a specific formation and avoiding trees. Such missions require adaptive path planning and inter-agent coordination to efficiently cover the area and mission success despite limited GPS access and visual obstructions \cite{c028a,c028b}. Similarly, coordinating multiple unmanned airships for long-duration surveillance or atmospheric data collection poses challenges including path tracking under wind disturbances, compensating for uncertain nonlinear dynamics, and maintaining safe distances between vehicles \cite{c029,c030}. These scenarios highlight the need for adaptive and decentralized control strategies capable of responding to both environmental and system-level uncertainties. To satisfy these requirements, formation control serves as a key approach. Its goal is to regulate how agents move relative to each other—their positions, velocities, and orientations—while maintaining a specific formation pattern \cite{c08}. It typically falls into two categories: centralized and distributed control. In centralized formation control, a single decision-making unit (central controller) has access to global information, including the states, dynamics, and goals of all agents in the network. The controller computes and transmits individual control signals to each agent to maintain the overall formation \cite{c041}. This approach benefits from simplified algorithm design, as the central controller typically has access to global state information and often assumes full knowledge of the system dynamics. For example, centralized strategies can exploit global optimization methods and perform sophisticated trajectory planning with minimal computational overhead per agent \cite{c042}. However, centralized control faces several well-known limitations. First, it introduces a single point of failure: if the central unit fails or communication is disrupted, the entire formation may collapse. Second, centralized schemes often lack scalability; as the number of agents increases, the computation and communication demands on the central controller grow rapidly, making real-time coordination difficult \cite{c043}. Third, centralized control requires high-bandwidth communication links to continuously transmit state and control information, which is often impractical in wireless, bandwidth-limited environments or over large spatial areas \cite{c044}.  In contrast, distributed formation control allows each agent to make decisions based on local information and, optionally, information from neighboring agents. Agents do not rely on global knowledge or a central controller; instead, they use local sensors and communication with neighbors to infer how to move in order to maintain the formation \cite{c045}. This decentralized decision-making enhances the scalability and robustness of the system. Distributed control is particularly advantageous in dynamic environments where agents need to adapt in real time to uncertainties such as disturbances, obstacles, and communication failures \cite{c046}.  Designing distributed control algorithms, however, is significantly more challenging. Without access to global state information, agents must rely on local estimations and consensus protocols to coordinate effectively. Ensuring convergence to the desired formation, stability under switching topologies or time-varying graphs, and robustness to noise and delays in communication are among the core technical issues \cite{c047, c048}.  Furthermore, achieving global objectives such as formation stabilization or trajectory tracking through only local interactions requires careful design using tools from graph theory, Lyapunov stability, and distributed optimization \cite{c049}. Recent research has addressed these challenges by incorporating adaptive and learning-based strategies into distributed formation control frameworks. For example, neural network-based distributed controllers have been proposed to approximate unknown nonlinear dynamics, enabling agents to both track desired formations and learn their dynamics online \cite{c0491}. Cooperative learning protocols further allow agents to share learned information, improving performance and accelerating convergence in uncertain environments \cite{c015}. These control architectures—centralized and distributed—can be applied across the three major formations: behavior-based, leader-following, and virtual structure approaches, each of which imposes different requirements and constraints on the information flow and coordination mechanisms among agents \cite{c05}.

First, behavioral approach, in which the formation problem is divided into several sub-problems, each with its own solution; and each agent is controlled by a predefined behavior derived as a weighted average of the sub-solutions \cite{c08,c09}. Second, the leader-following approach, in which agents follow the leader(s) while maintaining a specific configuration with them. Here, the overall formation behavior of the system is determined by designing a desired pattern for the leader(s). In this approach, since the followers are highly dependent on the leader(s)' information, they become vulnerable to the leader(s)' faults \cite{c010}. Finally, the virtual structure approach that is similar to the leader-following approach, with the key difference that agents follow a virtual leader. This makes the system more robust against leader's faults \cite{c011}.  In most of the leader-following formation control techniques the leader dynamics is linear time-invariant with no inputs \cite{c08, c013, c014}; while the system's ability to handle more complex formation tracking tasks requires the implementation of diverse and sophisticated reference trajectories. Yuan et al. at the cost of increased complexity and challenges, considered a more generalized leader dynamics model, where the leader model is driven by bounded and time-varying inputs and then designed a discontinuous distributed observer to estimate the leader’s state based solely on local communication~\cite{c015}. This advancement allowed agents to follow more expressive leader trajectories and stimulated learning along dynamic paths. However, their model assumes that the inertia matrix of each agent is known—a strong simplification, since in practical robotic systems, inertial properties often vary and are difficult to measure or model precisely. This limitation restricts the applicability of their approach in real-world conditions where the dynamics are completely uncertain, including time-varying inertia. Although some work has been done to account for unknown mass or inertia within adaptive control frameworks, these efforts typically still rely on simplified leader models and lack any cooperative mechanism across agents \cite{c054}. Thus, while existing methods have independently addressed either complex leader dynamics or dynamic uncertainties in agent models, the combined problem of achieving cooperative adaptive formation control in multi-agent systems with completely uncertain dynamics—including uncertain inertia—and non-zero input leader systems remains open. Solving this integrated challenge is essential for enabling scalable, adaptive, and intelligent formations in unstructured and dynamically evolving environments.

In this paper,  we propose a novel formation control scheme to address the formation control problem for a group of general nonlinear mechanical systems with complete uncertain dynamics under the virtual leader-following framework. The framework's control architecture consists of two layers. The first layer is a cooperative discontinuous nonlinear estimation protocol for estimating the leader's states, and the second layer introduces a novel cooperative formation control protocol, we propose in this work, to achieve cooperative formation tracking control for a group of agents with completely uncertain and nonlinear dynamics using neural networks (NNs). Compared to previous studies, our approach introduces three key innovations: (i) a more general system with an uncertain inertia matrix, (ii) an advanced formation protocol that enables cooperative online knowledge sharing by utilizing estimated NN weights from neighboring agents, enhancing knowledge consensus, and (iii) a more diverse reference trajectory incorporating time-varying external inputs for the leader.

\section{Methodology}

\subsection{Notation and Graph Theory }
The notations $I$, $1_n$, $\mathbb{R}$, $\mathbb{R}_+$,  $\mathbb{R}^n$ , $\mathbb{R}^{m \times n}$, $S^n$, $S_+^n$, $A \otimes B$ are used to represent the identity matrix,  the $n \times 1$  vectors  with all elements equal to $1$, the sets of real numbers, positive real numbers, real $n \times 1$ vectors, real $m \times n$ matrices, real symmetric $n \times n$ matrices, the symmetric positive definite matrices, and the Kronecker product of the two matrices $A$ and $B$, respectively. A block diagonal matrix with $X_1, X_2, \dots, X_p$ as its diagonal elements is expressed by $\mathrm{diag}\{X_1, X_2, \dots, X_p\}$. The notation $\mathrm{col}\{x_1, \ldots, x_n\}$ represents a column vector formed by stacking the column vectors $x_1, \ldots, x_n$ vertically.  The notations  $\mathbf{I}[k_1, k_2]$ and $\|x\| := (x^T x)^{1/2}$ denote the set of $\{k_1, k_1 + 1, \ldots, k_2\}$ for integers $k_1 < k_2$, and the norm of $x \in \mathbb{R}^n$, respectively.  For a square matrix $A$, $\lambda_i(A)$ is its $i^{th}$ eigenvalue, with $\overline{\lambda}(A)$ and $\underline{\lambda}(A)$ as the maximum and minimum eigenvalues.  A graph is defined by $\mathcal{G} = (\mathcal{V}, \mathcal{E}, \mathcal{A})$, where $\mathcal{V} = \{1, \dots, N\}$ is the set of vertices, $\mathcal{E}$ consists of edges $(i,j)$ with $i,j \in \mathcal{V}, i \neq j$, and $\mathcal{A} = [a_{ij}]_{N \times N}$ is the adjacency matrix. The Laplacian matrix of a  graph is given by $\mathcal{L} = [l_{ij}]$, where $l_{ii} = \sum_{j \neq i} a_{ij}$ and $l_{ij} = -a_{ij}$ for $i \neq j$.

\subsection{Radial Basis Function Neural Networks (RBF NNs)}
The RBF network with $N_n$ neurons is given by $f_{\text{nn}}(Z) = \sum_{i=1}^{N_n} w_i s_i(Z) = W^T S(Z)$ (see Fig. \ref{fig:topology} Right), where $s_i(\cdot)$ is the RBF, $Z \in \Omega_Z \subset \mathbb{R}^q$ is the input vector,  $W = [w_1, \dots, w_{N_n}]^T \in \mathbb{R}^{N_n}$ is the weight vector, and $S(Z) = [s_1(\|Z - \xi_1\|), \dots, s_{N_n}(\|Z - \xi_{N_n}\|)]^T$, with $\xi_i$ as distinct points in state space. The most common RBF is Gaussian function $s_i(\|Z - \xi_i\|) = \exp(-\frac{\|Z - \xi_i\|^2}{\gamma_i^2})$,  where $\xi_i = [\xi_{i1}, \dots, \xi_{iq}]^T$ and $\gamma_i$  are  the center and the the width of  the receptive field, respectively. Therefore, $f(Z) : \Omega_Z \to \mathbb{R}$ is a continuous function, so according to \cite{c016} if $\Omega_Z $ is a compact set, and $N_n $ is sufficiently large,  $f(Z) = W^{*T} S(Z) + \epsilon,$ where $W^*$ is an ideal constant weight vector and $|\epsilon| < \epsilon^*$ is the approximation error ($ \epsilon^* > 0$).\\
\textbf{Lemma 1:} For any continuous recurrent trajectory in a bounded compact set, the regressor subvector $S_\xi(Z)$, formed by $\xi_i$ near $Z(t)$, is persistently exciting (PE) \cite{c017}.

\subsection {Problem Statement}
\begin{figure}[t]
  \centering
  \includegraphics[width=0.38\textwidth, trim=175 390 253 256, clip]{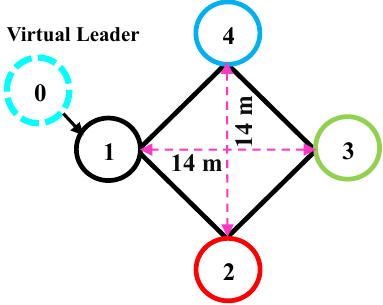}
  \hfill
 \includegraphics[width=0.58\textwidth, trim=95 450 70 50, clip]{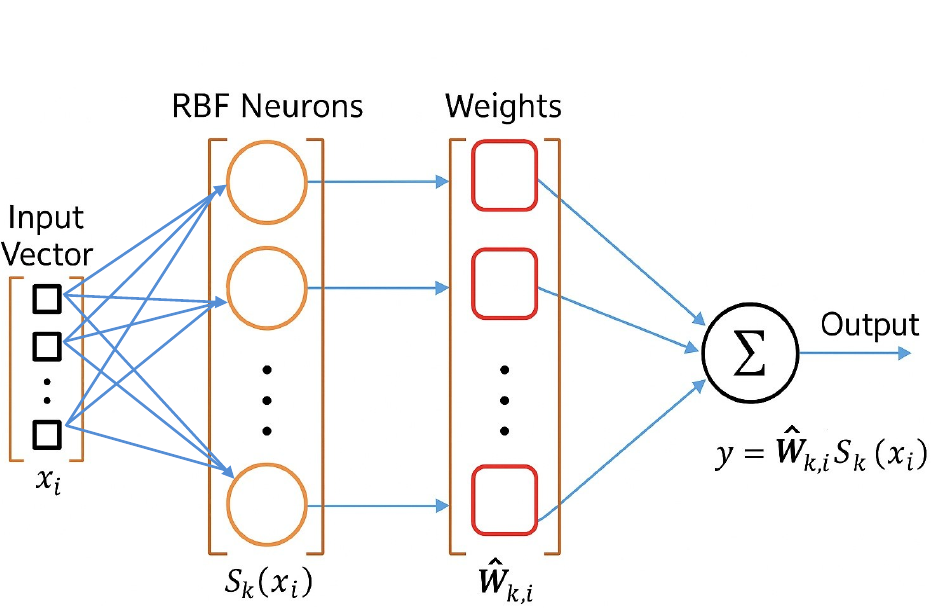}
  \hfill
  \caption{Left: Network topology and desired geometrical configuration. Right: RBF NNs structure. }
  \label{fig:topology}
\end{figure}

Consider a group of mechanical systems with nonlinear uncertain dynamics: 
\begin{align}
    &  \dot{p}_i = J(p_i) \nu_i \nonumber\\
    &  M\dot{\nu}_i + C(p_i, \nu_i) \nu_i + D(p_i, \nu_i) \nu_i + g(p_i) = \tau_i  \label{eq: 1}
\end{align}
where $i \in \mathbf{I}[1, N]$ with $N > 1$,   $p_i = (p_{i1}, \dots, p_{in})^T \in \mathbb{R}^n$, $v_i = (v_{i1}, \dots, v_{in})^T \in \mathbb{R}^n$,  $x_i := (p_i, v_i)^T \in \mathbb{R}^{2n} $, $\tau_i \in \mathbb{R}^n$, $M \in \mathbb{S}^n_+$ , $C(p_i, v_i) \in \mathbb{R}^{n\times n}$, $D(p_i, v_i) \in \mathbb{R}^{n\times n}$, $g(p_i) \in \mathbb{R}^{n\times 1}$ , and $J(p_i) \in \mathbb{R}^{n\times n}$, with $J(p_i) J^T(p_i) = I_n$ represent the number of the agents, position, velocity and state of the $i^{th}$ agent, the control input, an inertia matrix, the centripetal and Coriolis matrices, the friction terms, external forces due to gravity, and the rotation matrix, respectively. In this study we consider $ M, C(p_i, v_i), D(p_i, v_i)$ and $ g(p_i)$ as uncertain and unknown matrices.   We generate the desired formation trajectories by a virtual leader with bounded inputs: 
\begin{align}
\dot{p}_0 = v_0, \quad \dot{v}_0 = A_0 x_0 + B_0 r \label{eq:2}
\end{align}
where \( x_0 := (p_0, v_0)^T \in \mathbb{R}^{2n} \),  \( p_0 \in \mathbb{R}^n \),  \( v_0 \in \mathbb{R}^n \), \( r \in \mathbb{R}^{n_r} \),  \( A_0 \in \mathbb{R}^{n \times 2n}\) and \( B_0 \in \mathbb{R}^{n \times n_r} \)  standing for the leader’s state, position, velocity,  input signal,  and constant matrices, respectively. The network topology between mechanical systems and the leader is shown in Fig. \ref{fig:topology} Left, which is a fixed graph \( \mathcal{G} \). We have the following assumptions:\\
\textbf{Assumption 1}: The leader’s input signal is bounded ($\| r \| \leq r^*$,  where \( r^* \) is a positive constant).\\
\textbf{Assumption 2}: The leader's states are uniformly bounded, and the reference trajectory \( \varphi(x_0(t))|_{t \geq 0} \) is recurrent.\\
\textbf{Assumption 3}:  The leader has a directed path to at least one agent, and subgraph \( \mathcal{G}_s \) associated with \( N \) mechanical systems is undirected.

The Laplacian matrices of  \( \mathcal{G} \) and \( \mathcal{G}_s \) are \( \mathcal{L} \in \mathbb{R}^{(N +1) \times(N+1)} \) and \( \mathcal{L}_1 \in \mathbb{R}^{N \times N} \), respectively.  They are related by $\mathcal{L} =
\begin{bmatrix}
0 & 0_{1 \times N} \\
-\Delta \textbf{1}_{N } & \mathcal{L}_1
\end{bmatrix}$, where $\Delta = \text{diag} \{a_{10}, ..., a_{N0} \}$ with \( a_{i0} > 0 \) if the \( i^{th} \) agent is a neighbor of the leader, otherwise \( a_{i0} = 0 \). Under Assumption 3 $\mathcal{L}_1$ is symmetric and positive definite. We assume that only the leader's neighbors have access to its state information. Also, for distributed control design only $r^*$ is available, not  $r$. 

\begin{figure}[t]
  \centering
  \includegraphics[width=0.8\textwidth, trim=90 410 60 80, clip]{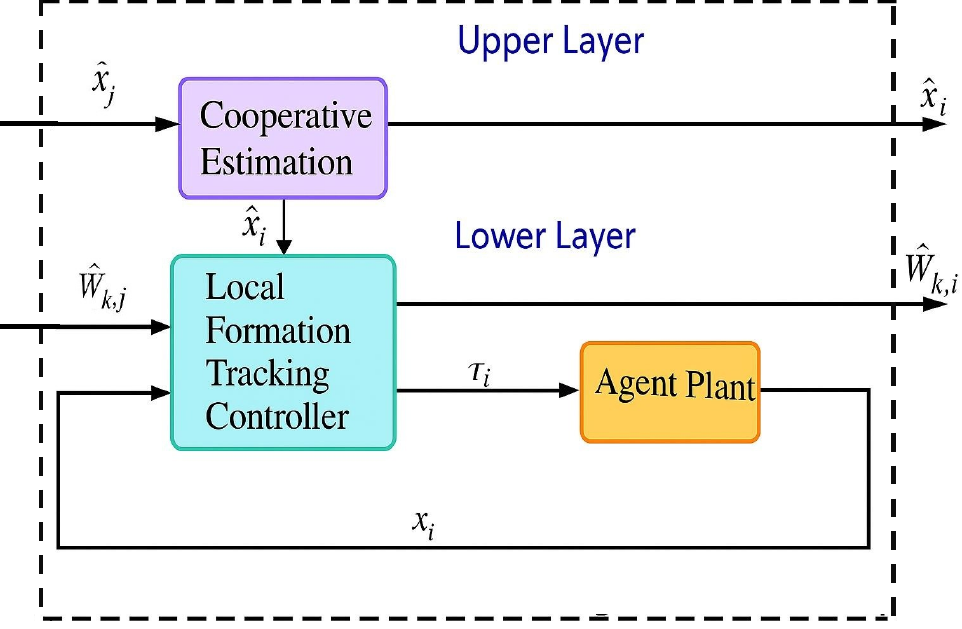}
  \hfill
  \caption{Proposed controller structure for each agent. The architecture consists of two hierarchical layers. In the upper layer, cooperative estimation is performed to estimate the virtual leader’s state using information from neighboring agents. The estimated state $\hat{x}_i$ serves as the reference input to the lower layer. In the lower layer, the local formation tracking controller uses the reference input along with estimated uncertainties $\hat{W}_{k,j}$ to compute the control input $\tau_i$, which is applied to the agent's plant. The agent’s actual state  $x_i$ is fed back to close the control loop, and the updated uncertainty estimate $\hat{W}_{k,i}$  is forwarded for network-wide adaptation.}
  \label{fig:twolayer}
\end{figure}

Here we aim to design a distributed RBF NN-based control protocol for the MAS (\ref{eq: 1})  under Assumptions 1$-$3 such that all agents track the leader while keeping a prespecified formation pattern.  To this end, we propose a novel formation control scheme consisting of an upper-layer cooperative estimator to estimate the leader’s state information, and a lower-layer cooperative deterministic learning control law to achieve the formation tracking control and cooperative learning (see Fig. \ref{fig:twolayer}).

%%%%%%%%%%%%%%%%%%%%%%%%%%%%%%%%%%%%%%%%%%
\subsection{Distributed Adaptive Control}

\subsubsection {First Layer: Cooperative Nonlinear Estimator}
We estimate the leader's states using the following cooperative discontinuous nonlinear estimation protocol: 

\begin{align}
\begin{bmatrix}
\dot{\hat{p}}_i \\ \dot{\hat{v}}_i
\end{bmatrix}
=
\begin{bmatrix}
0 & I_n \\
\multicolumn{2}{c}{A_0}
\end{bmatrix}
\begin{bmatrix}
\hat{p}_i \\ \hat{v}_i
\end{bmatrix}
+ \alpha_1 K_1 \phi_i 
+ \alpha_2 
\begin{bmatrix}
0 \\ B_0
\end{bmatrix}
f_1(K_2 \phi_i)
 \label{eq:3}
\end{align}

where $\hat{x}_i := (\hat{p}_i, \hat{v}_i)^T \in \mathbb{R}^{2n}$  is the observer state with $\hat{x}_0 = x_0$, $K_1 \in \mathbb{R}^{2n \times 2n}$, $K_2 \in \mathbb{R}^{n_r \times 2n}$, and $\alpha_1, \alpha_2 \in \mathbb{R}_+$ are observer coefficients, $
\phi_i = \sum_{j=0}^{N} a_{ij} (\hat{x}_i - \hat{x}_j)$, and, $f_1(K_2 \phi_i) =
\begin{cases} 
\frac{K_2 \phi_i}{\|K_2 \phi_i\|}, & \text{if } \|K_2 \phi_i\| \neq 0 \\
0, & \text{otherwise}
\end{cases}
$, ($\forall i \in \mathbf{I}[1,N]$). Defining $\tilde{x}_i = \hat{x}_i - x_0$ as the estimation error, it has been proved that $\tilde{x}_0 $ converges to zero exponentially, which means the boundedness of $\hat{x}_i := (\hat{p}_i, \hat{v}_i)^T$,  the exponential convergence of $\hat{p}_i \to p_0$, and $\hat{\nu}_i \to v_0 = \dot{p}_0 $ (Theorem 1 in \cite{c015}). Also, from \ref{eq:3}  $\dot{\hat{p}}_i = \hat{\nu}_i + \alpha_1 K_1 \phi_i $, since $\hat{x}_i \to x_0$ we have $\phi_i\to 0$, so $\dot{\hat{p}}_i \to \hat{\nu}_i \to  \dot{p}_0$.  The convergence proof and coefficient design ($K_1, K_2, \alpha_1, \alpha_2$) are detailed in \cite{c015}.

\subsubsection {Second Layer: Cooperative Deterministic Learning-Based Formation Control}
First we define $p_i^*$ as the constant desired vector between the $i^{th}$ agent’s position $p_i$ and the leader’s  $p_0$. Then, integrating backstepping adaptive control design \cite{c018} and deterministic learning using RBF NN \cite{c08} we design a formation control protocol such that $p_i$ track its reference trajectory: $p_{r,i} = p_0 + p_i^*$. To do so, as $p_0$ is not available for some agents, and because of the convergence of $\hat{p}_i \to p_0$, we use $\hat{p}_i$, a measurable signal, to define the reference trajectory $\hat{p}_{r,i} := \hat{p}_i + p_i^*$.  Considering the tracking error as $z_{1,i} = p_i - \hat{p}_{r,i}$, and knowing that \( \dot{\hat{p}}_{r,i}=\dot{\hat{p}}_i\) we have:
\begin{align}
     \dot{z}_{1,i} &= \dot{p_i}-\dot{\hat{p}}_{r,i} =  J(p_i) \nu_i, - \dot{\hat{p}}_i\label{eq:4}
\end{align}
Treating $\nu_i$ as a virtual control input (according to the backstepping technique \cite{c018}), we define a new variable as:
\begin{align}
     &\beta_i = {J}^T(p_i)(-H_{1,i}z_{1,i} + \dot{\hat{p}}_i) \label{eq: 5.5}\\
 &\dot{\beta}_i=  \dot{J}^T(p_i)(\dot{\hat{p}}_i-H_{1,i} z_{1,i}) + J^T(p_i)(\ddot{\hat{p}}_i-H_{1,i} \dot{z}_{1,i} ) \label{eq:8}\\
    & z_{2,i} = \nu_i - \beta_i=\nu_i-J(p_i)^T(-H_{1,i}z_{1,i} + \dot{\hat{p}}_i) \label{eq:6}\\
    &   \dot{z}_{2,i}=   \dot{\nu_i} - \dot{\beta_i} \label{eq:9}
\end{align}
where $H_{1,i}\in S_n^+$ is controller coefficient to be designed. Implementing (\ref{eq:6}) into (\ref{eq:4}) we have:
\begin{align}
    \dot{z}_{1,i}&=  -H_{1,i}z_{1,i} + J(p_i)z_{2,i} \label{eq:7} 
\end{align}
Substituting (\ref{eq:7}) into (\ref{eq:8}) and the result with (\ref{eq: 1}) into (\ref{eq:9}):
\begin{align}
    \dot{\beta}_i =& \dot{J}^T(p_i)(\dot{\hat{p}}_i-H_{1,i} z_{1,i} ) + J^T(p_i)H_{1,i}\hat{p}_i \nonumber\\
    &-  H_{1,i} v_i + J^T(p_i)\ddot{\hat{p}}_i \label{eq:9.5}\\
     \dot{z}_{2,i}= &- M^{-1} \left(C(p_i,\nu_i) \nu_i + D(p_i,\nu_i) \nu_i + g(p_i)  \right)  \nonumber \\
     & - \dot{\beta}_i  + M^{-1} \tau_i\label{eq:8.5}
\end{align}
We define $G(x_i)= M \dot{\beta}_i + C(p_i,\nu_i) \nu_i + D(p_i, \nu_i) \nu_i + g(p_i)=[G_1(x_i),...,G_n(x_i)]^T$  as a nonlinear function containing all the nonlinear uncertainty, and substitute it in (\ref{eq:8.5}):
\begin{align}
\dot{z}_{2,i} &=M^{-1}( \tau_i  - G(x_i)) \label{eq:10}
\end{align}
We employ RBF NNs to approximate the elements of $G(x_i)$:  
\begin{equation}
    G_k(x) = W_k^{*T} S_k(x) + \epsilon_k(x) \quad \forall k \in \mathbf{I}[1,n] \label{eq:11}
\end{equation}
where  $W_k^*$ is the ideal constant NN weight, and $\epsilon_k$ is the ideal approximation error satisfies  $|\epsilon_k| \leq \epsilon^*, \epsilon^*>0 $. Substituting (\ref{eq:11}) into (\ref{eq:10}): 
\begin{align}
\dot{z}_{2,i} =M^{-1}(\tau_i  - W^{*T} S(x_i) - \epsilon(x_i))  \label{eq:11.5}
\end{align}
where  $W^{*T} S(x_i) =  [W_1^{*T} S_1(x_i) ...W_n^{*T} S_n(x_i)]^T$ and $\epsilon (x_i)=  [ \epsilon_1(x_i)...\epsilon_n(x_i)]^T$. Since $W_k^*$ for all $k \in \mathbf{I}[1,n]$ is unknown, we utilize $\hat{W}_{k,i}$ to estimate $W_k^*$ for the $i^{th}$ agent. Then, we propose the following RBF NN-based cooperative deterministic learning:
\begin{align}
   \tau_i= & \hat{W}_i^T S^F(x_i) -H_{2,i}z_{2,i}-J^T(p_i)z_{1,i} \label{eq:13}\\
     \dot{\hat{W}}_{k,i} = &-\gamma_1( S_k(x_i) z_{2,i} + \sigma \hat{W}_{k,i})  \nonumber \\
     & - \gamma_2 \sum_{j=1}^N a_{ij} ( \hat{W}_{k,i} - \hat{W}_{k,j} ) \label{eq:14}
\end{align}             
where  $ \hat{W}_i^T S^F(x_i) := \begin{bmatrix} \hat{W}_{1,i}^T S_1(x_i) \cdots \hat{W}_{n,i}^T S_n(x_i) \end{bmatrix}^T \nonumber$, and $H_{2,i} \in S_n^+$, $\gamma_1, \gamma_2$, and $\sigma \in \mathbb{R}_+$ are controller coefficients to be designed (for $\forall i \in \mathbf{I}[1,N], k \in \mathbf{I}[1,n]$). We consider $\tilde{W}_{k,i} := \hat{W}_{k,i} - W_k^*$ , interconnecting (\ref{eq:13}) and (\ref{eq:11.5}) we formulate the closed-loop system for each agent in terms of the variables $z_{1,i}$, $z_{2,i}$ and $\tilde{W}_{k,i}$ as follows:
\begin{align} 
   \dot{z}_{1,i}=&  -H_{1,i}z_{1,i} + J(p_i)z_{2,i} \nonumber\\
   \dot{z}_{2,i}=&M^{-1}(\tilde{W}_{i}^TS(x_i)-J^T(p_i)z_{1,i}-H_{2,i}z_{2,i}-\epsilon(x_i))\nonumber\\
   \dot{\tilde{W}}_{k,i}=&-\gamma_1 \left( S_k(x_i) z_{2,i} + \sigma \hat{W}_{k,i} \right)  \nonumber\\
   &- \gamma_2 \sum_{j=1}^N a_{ij} \left( \tilde{W}_{k,i} - \tilde{W}_{k,j} \right)\label{eq:15}
   \end{align} 
where $\tilde{W}_i^T S^F(x_i) := \begin{bmatrix} \tilde{W}_{1,i}^T S_1(x_i) \cdots \tilde{W}_{n,i}^T S_n(x_i) \end{bmatrix}^T$, $z_{2,i} = \begin{bmatrix} z_{2,1,i} \cdots z_{2,n,i} \end{bmatrix}^T$, and $\epsilon(x_i) := \begin{bmatrix} \epsilon_1(x_i) \cdots \epsilon_n(x_i) \end{bmatrix}^T$ (for $\forall i \in \mathbf{I}[1,N], k \in \mathbf{I}[1,n]$). (Note that \(\dot{W}^{*}_{k}=0\), so $ \dot{\tilde{W}}_{k,j}=\dot{\hat{W}}_{k,j}$, and $\hat{W}_{k,i} - \hat{W}_{k,j}=\tilde{W}_{k,i}-\tilde{W}_{k,j}$)\\

\textbf{Theorem 1:} Consider the MAS (\ref{eq: 1}) and the reference model (\ref{eq:2}) with the cooperative estimation law (\ref{eq:3}), the cooperative formation learning control law (\ref{eq:13}), and the cooperative NN weight updating law (\ref{eq:14}). Assume there exists a sufficiently large compact set $\Omega_i$ in which the NN input vectors remain ($x_i \in \Omega_i$), and the tracking trajectory of the $i^{th}$ mechanical system $\varphi_{\zeta_i}$ is recurrent and remains in a bounded compact set (e.g. $\varphi_{\zeta_i}\in \Omega_i$). Also assume all initial conditions are bounded, $\hat{W}_{k,i}(0) = 0$ ($\forall t\geq 0$, $\forall i\in \mathbf{I}[1,N]$ , and $\forall k\in \mathbf{I}[1,n]$), and the control coefficients are chosen properly (i.e.,  $\underline{\lambda}(H_{1,i}), \underline{\lambda}(H_{2,i})$ are sufficiently large,  $ \sigma > 0$  is sufficiently small, $ \gamma_1 > 0$, and $\gamma_2 > 0$, we have: 

\begin{enumerate}
    \renewcommand{\labelenumi}{\textit{\roman{enumi})}}
    \item \textbf{Stability}: all signals in the closed-loop system remain uniformly ultimately bounded (UUB);
    \item \textbf{Tracking}: the position tracking error $p_i - p_{r,i}$ converges exponentially to a small neighborhood of zero.
\end{enumerate}

\section{Results}
We begin the results section by theoretically proving that the proposed cooperative formation control scheme ensures system stability and guarantees convergence of the tracking error  to a small neighborhood of zero. This theoretical result is further validated through a numerical example implemented in MATLAB.

\subsection{Theoretical Analysis }
\subsubsection{ Proof of Stability}
Consider the following Lyapunov function candidate for the closed-loop system (\ref{eq:15}):
\begin{align}
    V = \sum_{i=1}^N(\frac{1}{2} z_{1,i}^T z_{1,i} + \frac{1}{2} z_{2,i}^T M z_{2,i} + \sum_{k=1}^n \frac{\tilde{W}_{k,i}^T \tilde{W}_{k,i}}{2\gamma_{1}})\label{eq: 17}
\end{align}
Derivative of $V$ yields (Note that here, the inertia matrix is considered unknown but constant, $ \dot{M}=0$):
\begin{align}
    \dot{V}  =& \sum_{i=1}^N \left(z_{1,i}^T\dot{z}_{1,i}  + z_{2,i}^TM\dot{z}_{2,i} + \sum_{k=1}^{n} \tilde{W}_{k,i}^T \gamma_{1}^{-1} \dot{\tilde{W}}_{k,i}\right)= \nonumber\\
&-\sum_{i=1}^{N}\textbf{\LARGE(} z_{1,i}^T H_{1,i} z_{1,i} +  z_{2,i}^T H_{2,i} z_{2,i} + z_{2,i}^T\epsilon(x_i) \textbf{\LARGE) }  \nonumber \\
&  - \sum_{i=1}^{N}\sum_{k=1}^{n}\tilde{W}_{k,i}^T \sigma \hat{W}_{k,i} -\frac{\gamma_2}{\gamma_1} \sum_{k=1}^{n} \mathbf{\tilde{W}}_k^T ( \mathcal{L}_1 \otimes I )\mathbf{\tilde{W}}_k \nonumber
\end{align}
where $\mathbf{\tilde{W}}_k := \operatorname{col}\{\tilde{W}_{k,1}, \dots, \tilde{W}_{k,N} \}, (\forall k \in \mathbf{I}[1,n])$ and $\mathcal{L}_1$ is the Laplacian matrix of the network graph associated with the follower agents \( \mathcal{G}_s \), which is undirected (Fig. \ref{fig:topology} Left); so, $\mathcal{L}_1$ is positive semidefinite real symmetric matrix ($\underline{\lambda}(\mathcal{L}_1) = 0$), results in \(\frac{\gamma_2}{\gamma_1} \sum_{k=1}^{n} \mathbf{\tilde{W}}_k^T ( \mathcal{L}_1 \otimes I )\mathbf{\tilde{W}}_k\geq0\). We choose $H_{2,i} = H_{1,i} + H_{22,i}$ such that $H_{1,i}\in \mathbb{S}^n_+$, $ H_{22,i} \in \mathbb{S}^n_+$, $\underline{\lambda}(H_1,i)\geq0 $, and $\underline{\lambda}(H_{22},i)\geq0 $. Thus, we have: 
\begin{align}
\dot{V}\leq&- \sum_{i=1}^{N} ( z_{1,i}^T H_{1,i} z_{1,i}+ z_{2,i}^T H_{1,i} z_{2,i} + z_{2,i}^T H_{22,i} z_{2,i}) \nonumber\\
&+\sum_{i=1}^{N}z_{2,i}^T \epsilon(x_i) +\sum_{i=1}^{N}\sum_{k=1}^{n}\tilde{W}_{k,i}^T \sigma \hat{W}_{k,i} \label{eq: 17.5}
\end{align}
\\
Since $H_{1,i}\in S_n^+$, we have $z_{1,i}^T H_{1,i} z_{1,i}\geq \underline{\lambda}(H_1,i)z_{1,i}^Tz_{1,i}$. So, $-z_{1,i}^T H_{1,i} z_{1,i}\leq-2\underline{\lambda}(H_{1,i})(\frac{1}{2}z_{1,i}^Tz_{1,i})\leq0$. in the same manner,  $-z_{2,i}^Tz_{2,i}\leq\frac{-2}{\overline{\lambda}(M)}(\frac{1}{2}z_{2,i}^TMz_{2,i})$, as $M\in S_n^+$,  and $\overline{\lambda}(M)>0$. Also, applying Cauchy-Schwarz inequality to $-\tilde{W}_{k,i}^T \sigma \hat{W}_{k,i}=-\sigma||\tilde{W}_{k,i}||^2  -\sigma\tilde{W}_{k,i}^T  W^{*}_{k}$ results in $ -\sigma\tilde{W}_{k,i}^T  W^{*}_{k} \leq \sigma||\tilde{W}_{k,i}|| ||W^{*}_{k}||$. Furthermore, applying completion of squares inequality results in $||\tilde{W}_{k,i}|| ||W^{*}_{k}||\leq \frac{1}{2}||\tilde{W}_{k,i}||^2+ \frac{1}{2}||W^{*}_{k}||^2$. Combining the two results, we obtain: $-\sigma||\tilde{W}_{k,i}|| ||W^{*}_{k,i}||\leq \frac{\sigma}{2}(||\tilde{W}_{k,i}||^2+||W^{*}_{k}||^2 )$.  Therefore, we obtain $-\tilde{W}_{k,i}^T \sigma \hat{W}_{k,i} \leq -  \frac{\sigma}{2}\tilde{W}_{k,i}^T\tilde{W}_{k,i}+ \frac{\sigma}{2}||W^{*}_{k}||^2$. Also, we have $-z_{2,i}^T H_{22,i} z_{2,i}  \leq  -\underline{\lambda}(H_{22,i})||z_{2,i}||^2 $ because $H_{22,i}\in S_n^+$. Moreover, $-z_{2,i}^T \epsilon(x_i)\leq ||z_{2,i}^T||||\epsilon(x_i)||$. It follows that, $-z_{2,i}^T H_{22,i} z_{2,i} -z_{2,i}^T \epsilon(x_i)\leq  -\underline{\lambda}(H_{22,i})||z_{2,i}||^2 + ||z_{2,i}^T||||\epsilon(x_i)|| =  -\underline{\lambda}(H_{22},i)(z_{2,i} - \frac{\epsilon(x_i)}{2\underline{\lambda}(H_{22,i})})^2 + \frac{\epsilon(x_i)^T\epsilon(x_i)}{4\underline{\lambda}(H_{22,i})}$. As a result, since $ |\epsilon(x_i)| < \epsilon^*$, we achieve  $ -z_{2,i}^T H_{22,i} z_{2,i} -z_{2,i}^T \epsilon(x_i)\leq  \frac{||\epsilon^*||^2}{4\underline{\lambda}(H_{22},i)}$ . Integrating all the derived inequalities we conclude: 
\begin{align}
    \dot{V}&\leq  \sum_{i=1}^{N} \textbf{\Large (} -2\underline{\lambda}(H_{1,i})(\frac{1}{2}z_{1,i}^Tz_{1,i}) -\frac{2\underline{\lambda}(H_{1,i})}{\overline{\lambda}(M)}(\frac{1}{2}z_{2,i}^TMz_{2,i}) \nonumber\\
    &-\sigma \gamma_1\sum_{k=1}^{n} (\frac{\tilde{W}_{k,i}^T\tilde{W}_{k,i}}{2\gamma_1})+ 
\frac{||\epsilon^*||^2}{4\underline{\lambda}(H_{22,i})}+\sum_{k=1}^{n}\frac{\sigma }{2}||W^{*}_{k,i}||^2\textbf{\Large )}  \nonumber\\
& \dot{V}\leq -\rho V+ \delta \label{eq: 18}
\end{align}
where $\delta := \sum_{i=1}^{N}  (
\frac{||\epsilon^*||^2}{4\underline{\lambda}(H_{22},i)}+\sum_{k=1}^{n}\frac{\sigma }{2}||W^{*}_{k,i}||^2) $ and $\rho := \text{min} \{2\underline{\lambda}(H_{1,i}), \frac{2\underline{\lambda}(H_{1,i})}{\overline{\lambda}(M)}, \sigma \gamma_1\} $. Given that $0 < \delta < \infty$, equation (\ref{eq: 18}) ensures the UUB of $z_{1,i}$, $z_{2,i}$, and $\tilde{W}_{k,i}$. Due to the boundedness of  $\hat{p}_i $ \cite{c015},  $p_{r,i} = \hat{p}_i + p_i^*$ is also bounded thereby ensuring the boundedness of  $p_i = z_{1,i} + \hat{p}_{r,i}$. Similarly, due to the boundedness of $J^T(p_i)$, $z_{1,i}$ and $\dot{\hat{p}}_i$ (Theorem 1 in \cite{c015} and Assumption 2) $\beta_i$ (\ref{eq: 5.5}) is also bounded, which ensures boundedness of $\nu_i = z_{2,i} + \beta_i$.  Moreover, boundedness of  $\tilde{W}$ and $W^*$ leads to the boundedness of $\hat{W}_{k,i} = \tilde {W}_{k,i} + W_k^*$.  Thus,  the control signal $\tau_i$ (\ref{eq:13}) is also bounded because of the boundedness of $z_{1,i}$, $z_{2,i}, J(p_i)$, $\hat{W}_{k,i}$ , and the Gaussian function regressor $S^F(x_i)$ \cite{c017} ($\forall i \in \mathbf{I}[1,N], k \in \mathbf{I}[1,n]$). As a result, all the signals in the closed-loop system remain UUB (proof of stability in Theorem 1). 

\subsubsection{Proof of Tracking}
Consider the following Lyapunov function candidate for the dynamics of $z_{1,i}$ and $z_{2,i}$ in (\ref{eq:15}):
\begin{align}
    V_z = \sum_{i=1}^N\frac{1}{2} z_{1,i}^T z_{1,i} + \sum_{i=1}^N\frac{1}{2} z_{2,i}^T M z_{2,i} \label{eq:19}
\end{align}
Derivative of $V_z$ yields:
\begin{flalign*}
    \dot{V}_z = \sum_{i=1}^N& \textbf{\Large (} z_{2,i}^T(\tilde{W}_{i}^T S^F(x_i)-H_{2,i} z_{2,i}  -\epsilon(x_i))- z_{1,i}^TH_{1,i} z_{1,i}\textbf{\Large )}
\end{flalign*} 

To proceed we choose $H_{2,i}= H_{1,i}+2H_{22,i}$ such that  $H_{22,i}\in \mathbb {S^n_+}$, which results in $\dot{V}_z = \sum_{i=1}^N \textbf{\Large (}(-z_{1,i}^TH_{1,i} z_{1,i} -  z_{2,i}^TH_{1,i} z_{2,i}) +  (- z_{2,i}^TH_{22,i} z_{2,i}+ z_{2,i}^T\tilde{W}_{i}^T S^F(x_i)) +(-z_{2,i}^TH_{22,i} z_{2,i} - z_{2,i}^T\epsilon(x_i))\textbf{\Large )}$. Similar to the stability proof, given that the Gaussian function regressor $S^F(x_i)$ is bounded \cite{c017} ($\|S^F(x_i)\| \leq s^*,  0 < s^* < \infty
, \forall i \in \mathbf{I}[1,N]$) and $H_{22,i}\in \mathbb {S^n_+}$ we apply the Cauchy-Schwarz along with the completion of squares inequalities to obtain $- z_{2,i}^TH_{22,i} z_{2,i}+ z_{2,i}^T\tilde{W}_{i}^T S^F(x_i) \leq \frac{||\tilde{W}_{i}^*||^2 s^{*2}}{4\underline{\lambda}(H_{22},i)}$,  where $\tilde{W}_i^* = 
\begin{bmatrix}
\tilde{W}_{1,i}^*  \cdots  \tilde{W}_{n,i}^*
\end{bmatrix}^T
\text{ with } \tilde{W}_{k,i}^* $ denoting the upper bound of $\tilde{W}_{k,i}$ ($\forall i \in \mathbf{I}[1,N], k \in \mathbf{I}[1,n]$) that exists due to its boundedness proved in (i). Likewise, $-z_{2,i}^T H_{22,i} z_{2,i} -z_{2,i}^T \epsilon(x_i)\leq  \frac{||\epsilon^*||^2}{4\underline{\lambda}(H_{22},i)}$. Integrating the derived inequalities leads to the conclusion that:
\begin{align}
    &\dot{V}_z\leq \sum_{i=1}^{N} \textbf{\Large (} -2\underline{\lambda}(H_{1,i})(\frac{1}{2}z_{1,i}^Tz_{1,i}) -\frac{2\underline{\lambda}(H_{1,i})}{\overline{\lambda}(M)}(\frac{1}{2}z_{2,i}^TMz_{2,i}) \nonumber\\
    & +\frac{||\epsilon^*||^2}{4\underline{\lambda}(H_{22,i})}+\frac{||\tilde{W}_{i}^*||^2 s^{*2}}{4\underline{\lambda}(H_{22,i})}\textbf{\Large )} \Rightarrow \dot{V}_z\leq-\rho_z V_z+ \delta_z \label{eq: 20}
\end{align}
where $\rho_z := \text{min} \{2\underline{\lambda}(H_{1,i}), \frac{2\underline{\lambda}(H_{1,i})}{\overline{\lambda}(M)}\}$ and $\delta_z := \sum_{i=1}^{N}  (
\frac{||\epsilon^*||^2}{4\underline{\lambda}(H_{22},i)}+\frac{||\tilde{W}_{i}^*||^2 S^{*2}}{4\underline{\lambda}(H_{22},i)}) > 0$. From (\ref{eq:19}) and (\ref{eq: 20})  we have $\text{min }\{1,\underline{\lambda}(M) \}(\frac{1}{2}\sum_{i=1}^{N}(||z_{1,i}||^2+||z_{2,i}||^2))\leq V_z\leq V_z(0)\exp{(-\rho_z t)} + \frac{\delta_z}{\rho_z}$, so: \\
\begin{align}
\sum_{i=1}^{N}(||z_{1,i}||^2+||z_{2,i}||^2)\leq \frac{2}{\text{min} \{1,\underline{\lambda}(M) \}}V_z(0)\exp{(-\rho_z t)} + \frac{2\delta_z}{\rho_z \text{min} \{1,\underline{\lambda}(M) \}}
\end{align}
We consider  $ \bar{\delta}_z$ as an upper band for $ \sqrt{\frac{2\delta_z}{\rho_z \text{min} \{1,\underline{\lambda}(M) \}}} < \overline{\delta_z}$.  So, there exist a finite time $ T>0$ such that for all $t>T$, $||z_{1,i}|| \leq   \overline{\delta_z}$ and $||z_{2,i}|| \leq   \overline{\delta_z}$. By choosing sufficiently large $\underline{\lambda}(H_{1,i}) > 0$ and 
$\underline{\lambda}(H_{2,i}) > 0$ ($\forall i \in \mathbf{I}[1,N]$)  $\delta_z$ can be taken arbitrary small, which leads to the exponential convergence of $z_{1,i}$ and $z_{2,i}$  to a small neighborhood of zero, associate with  exponential convergence of $p_i \rightarrow \hat{p}_{r,i}$ and $\nu_i \rightarrow \beta_i$   respectively.  Therefore, due to the exponential convergence of $\hat{p}_i \to p_0$ \cite{c015}, $\hat{p}_{r,i}$ will converge to a small neighborhood of $  p_{r,i}$ exponentially, results in the exponential convergence of $ p_i$ to a small neighborhood of  $p_{r,i} = p_0 + p_i^*$ (proof of tracking in Theorem 1).

\begin{figure}[t]
  \centering
  \subfloat[$\hat{p}_{i1} \rightarrow p_{01}$]{\includegraphics[width=0.33\textwidth, trim=80 210 70 200, clip]{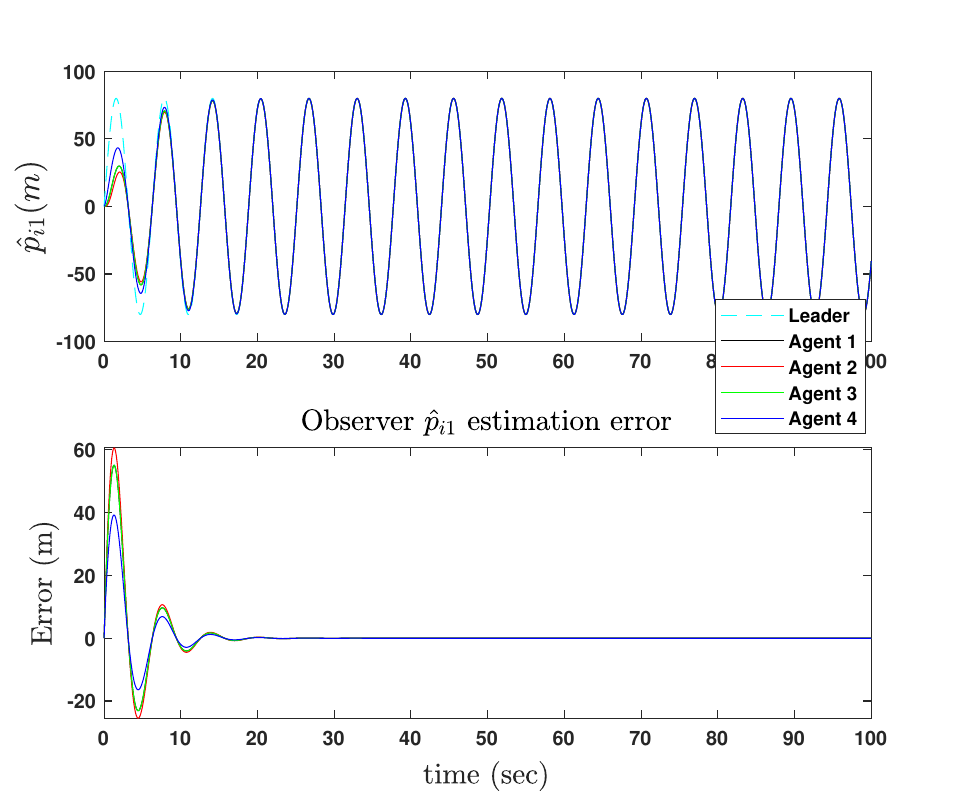}}
  \hfill
  \subfloat[$\hat{p}_{i2} \rightarrow p_{02}$]{\includegraphics[width=0.33\textwidth, trim=80 210 70 200, clip]{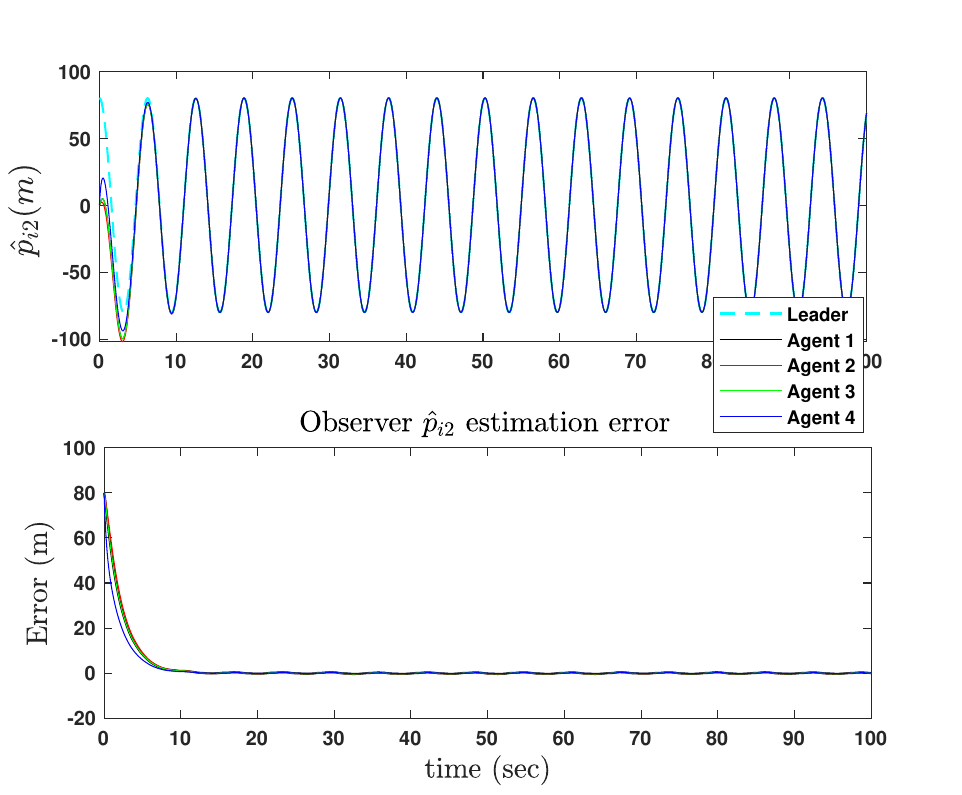}}
  \hfill
  \subfloat[$\hat{p}_{i3} \rightarrow p_{03}$]{\includegraphics[width=0.33\textwidth, trim=80 210 70 200, clip]{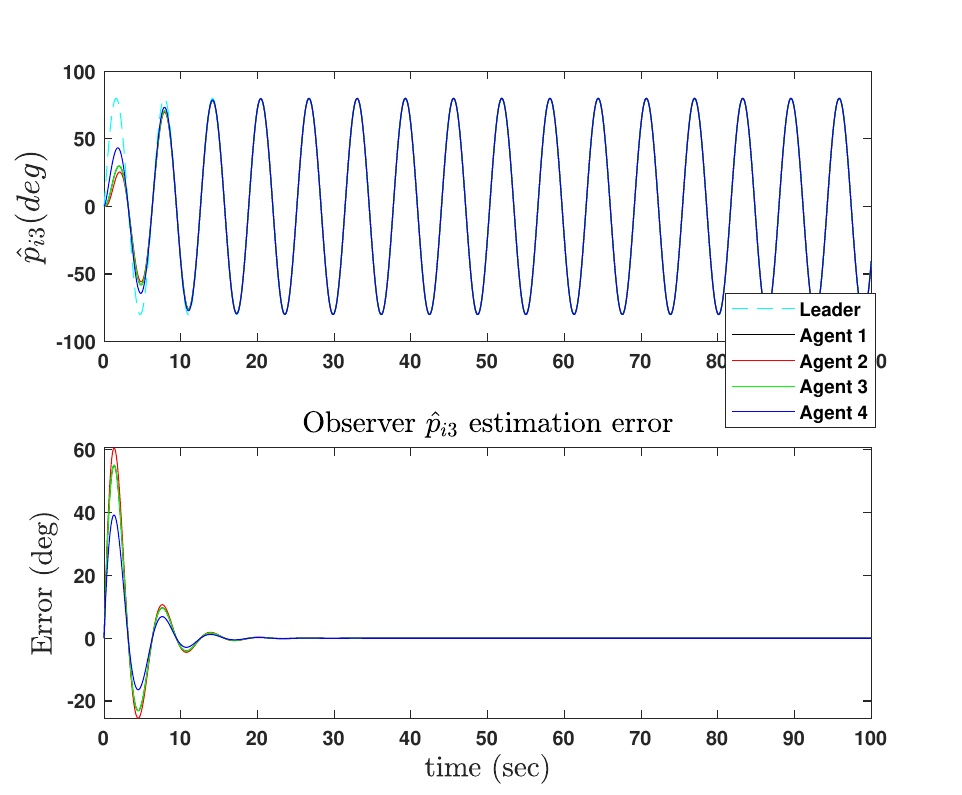}}
  \caption{Cooperative nonlinear estimation. The top row shows the estimated trajectory by each agent, while the bottom row illustrates the corresponding estimation errors.}
  \label{fig:estimator}
\end{figure}

\begin{figure}[t]
  \centering
  \subfloat[${p}_{i1} \rightarrow p_{01}$]{\includegraphics[width=0.33\textwidth, trim=80 210 70 200, clip]{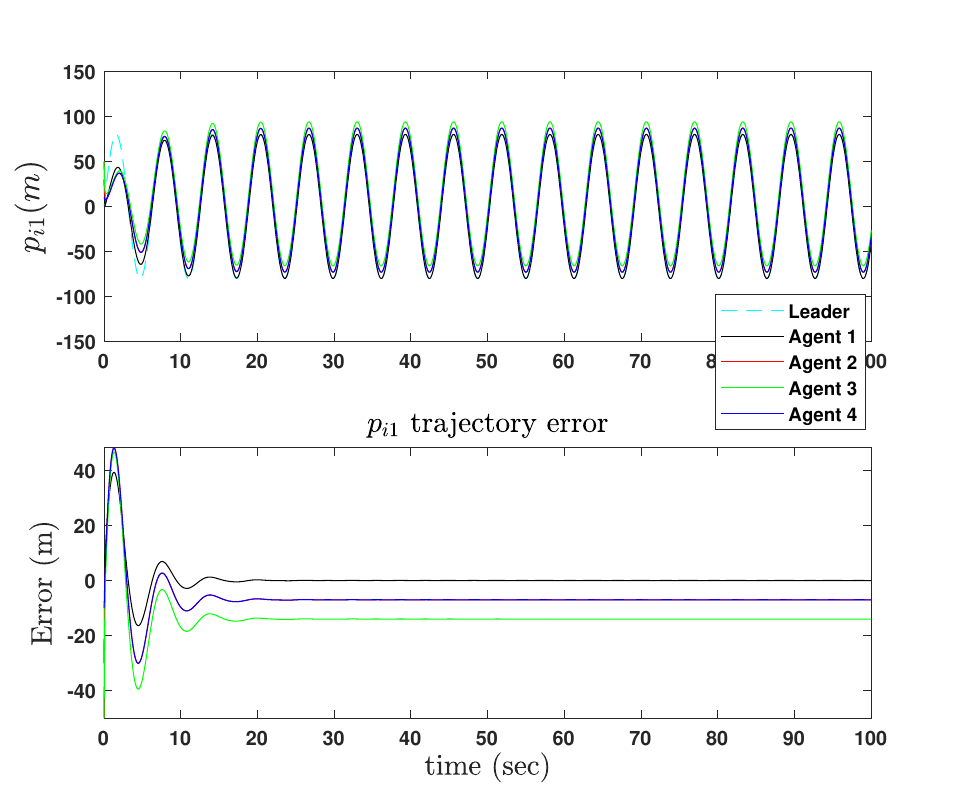}}
  \hfill
  \subfloat[${p}_{i2} \rightarrow p_{02}$]{\includegraphics[width=0.33\textwidth, trim=80 210 70 200, clip]{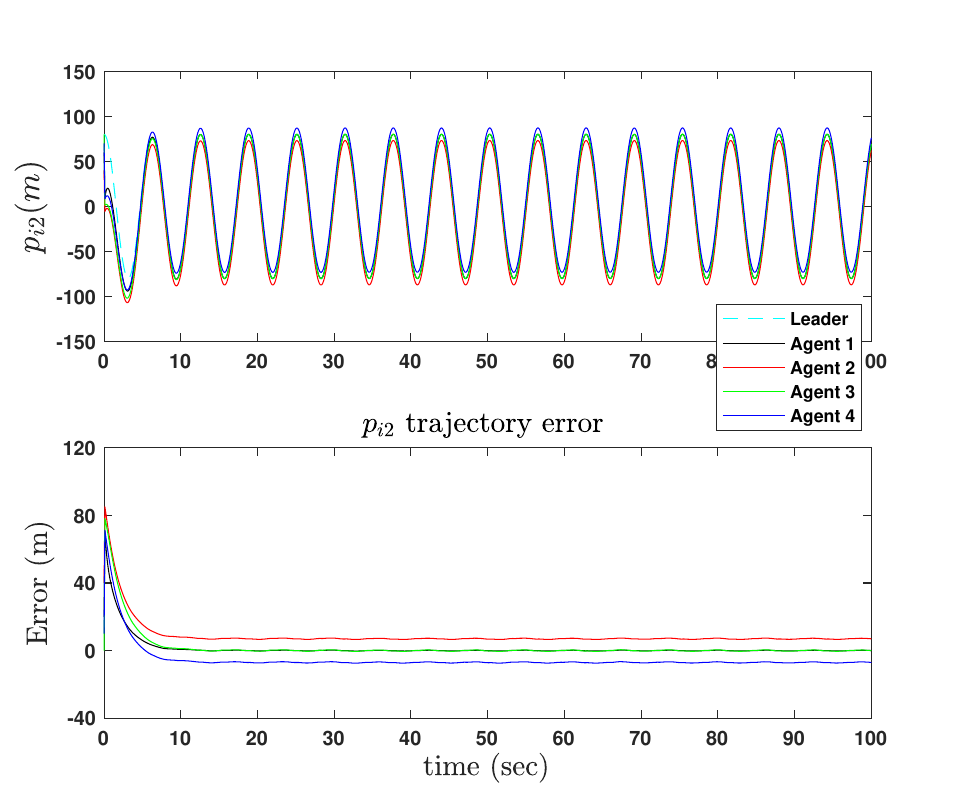}}
  \hfill
  \subfloat[${p}_{i3} \rightarrow p_{03}$]{\includegraphics[width=0.33\textwidth, trim=80 210 70 200, clip]{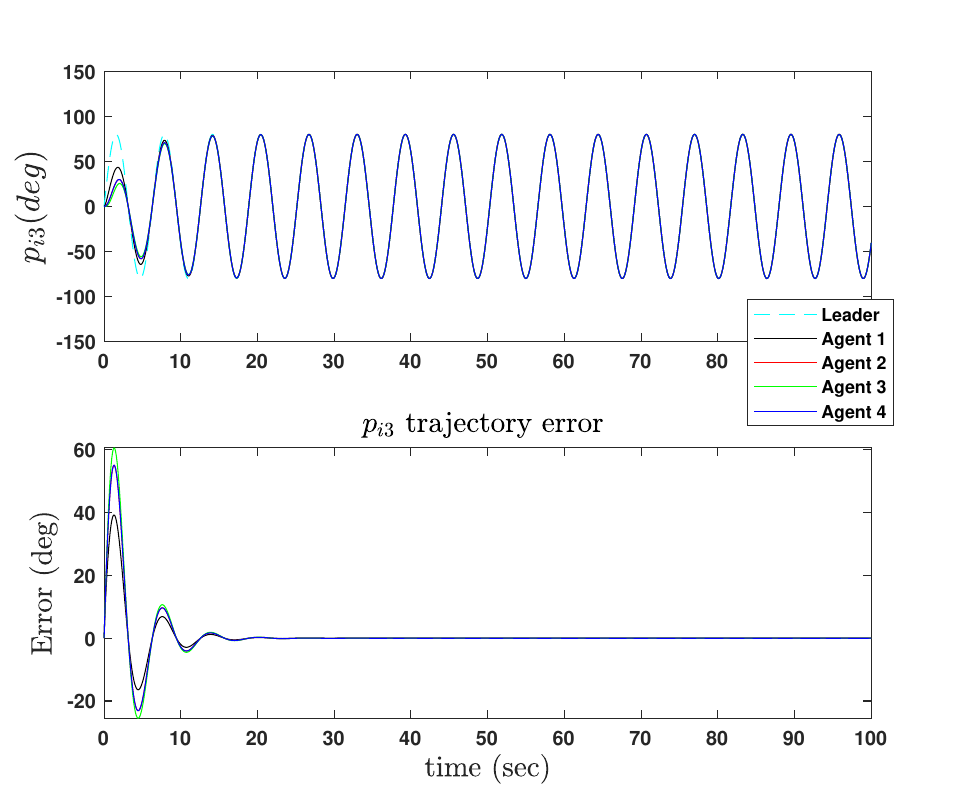}}
  \caption{Position tracking responses in the $p_{i1}$, $p_{i2}$, and $p_{i3}$ dimensions. The top row shows the actual trajectories of the agents and the leader in each dimension, while the bottom row illustrates the corresponding tracking errors.}
  \label{fig:positiontracking}
\end{figure}

\subsection{Experimental Verification }

We consider a group of four MASs \cite{c08} in the form of (\ref{eq: 1}) with $M =\Bigl[
\begin{smallmatrix}
25 & 0 & 0 \\
0 & 33 & 1.15 \\
0 & 1.15 & 2.8
\end{smallmatrix}\Bigr]$, $D(v_i) =
\bigl[\begin{smallmatrix}
0.8 + 1.3|v_{i1}| & 0 & 0 \\
0 & 0.9 + 36|v_{i2}| & -0.1 \\
0 & -0.1 & 0
\end{smallmatrix}\bigr]$,  $C(v_i) =
\Bigl[\begin{smallmatrix}
0 & 0 & -33v_{i2} - 1.15v_{i3} \\
0 & 0 & 25v_{i1} \\
33v_{i2} + 1.15v_{i3} & -25v_{i1} & 0
\end{smallmatrix}\Bigr]$, $g(p_i) = 0$, and $J(p_i)=I_3$. All the system parameters, including the inertia matrix \( M \), damping matrix \( D(v_i) \), Coriolis matrix \( C(v_i) \), and gravitational term \( g(p_i) \), are assumed to be completely uncertain. In this study, we define the time-varying input of the virtual leader as \( r = -80\cos(t) \),  generating a smooth time-varying reference trajectory to simulate a realistic mission. This dynamic input evaluates the proposed control scheme’s capability to track the desired trajectory while maintaining the specific formation configuration under continuously varying conditions. For future research, we aim to integrate brain signals as inputs to the reference trajectory, allowing human intention to guide the system’s desired motion. For details on interpreting human brain signals, refer to \cite{c022}, \cite{c024}. Also, we consider $
A_0 =
\bigl[\begin{smallmatrix}
-1 & 0 & 0 & 0 & 0 & 0  \\
0 & 0 & 0 & 0 & 0 & 0  \\
0 & 0 & -1 & 0 & 0 & 0 
\end{smallmatrix}\bigr]$, $B_0 =[0, 1, 0]^T$for the virtual leader dynamics (\ref{eq:2}) with initial condition  $p_0(0) = [0, 80, 0]^T$, $v_0(0) = [80, 0, 80]^T$. Using an observer similar to that of \cite{c015}, with $\alpha_1 = 1$ and  $\alpha_2 = 200$ and  zero initial condition, we can perfectly estimate the leader's state in the presence of bounded unknown leader inputs. As shown in Fig. \ref{fig:estimator}, the estimation errors for all agents exhibit fast exponential convergence to zero, confirming the effectiveness and robustness of the observer design under uncertain and time-varying leader dynamics.

\begin{figure}[t]
    \centering
        \includegraphics[width=0.6\textwidth, trim=0 90 0 80, clip]{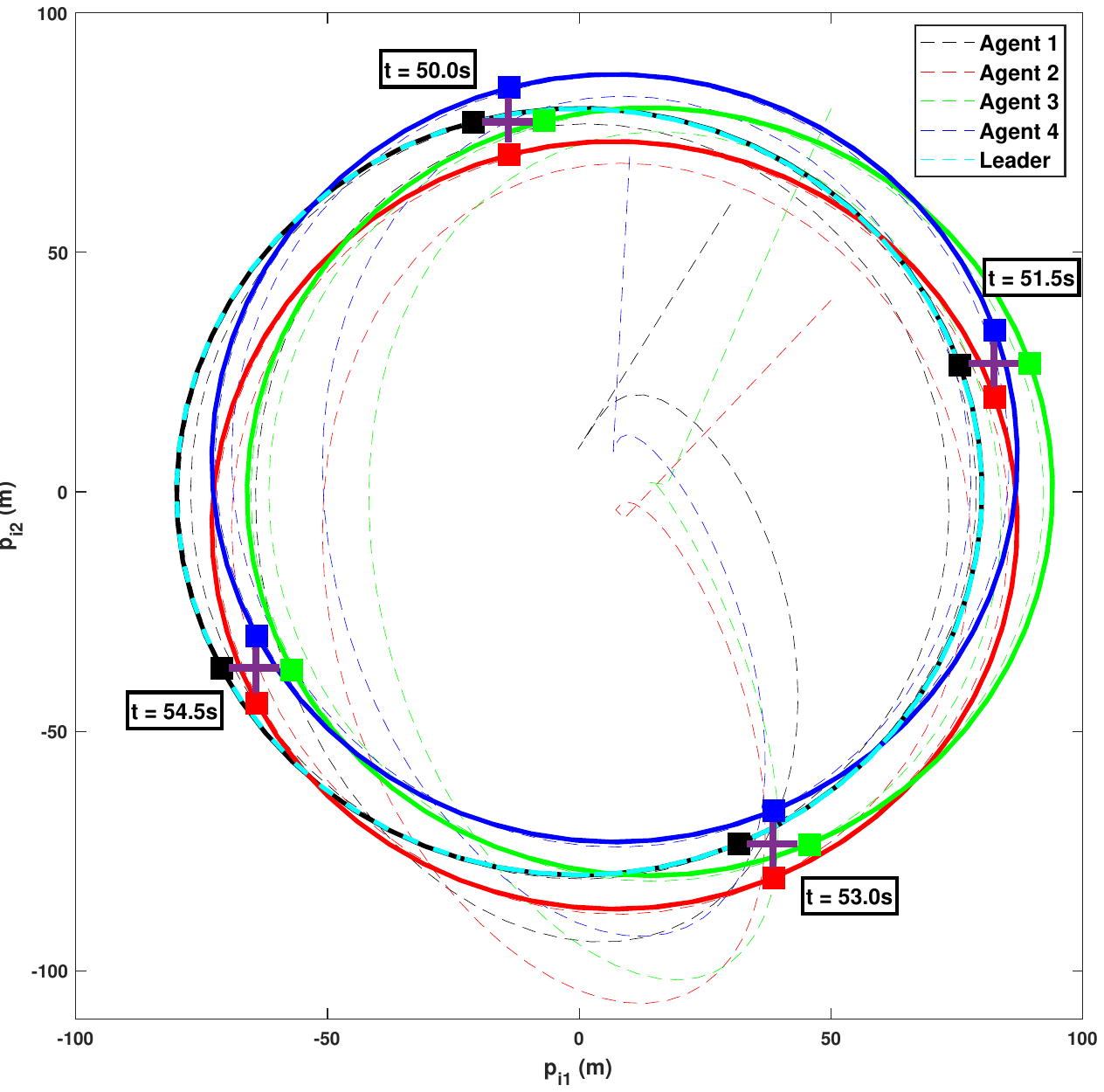}
    \caption{Formation tracking of the MAS in the $p_{i1}$–$p_{i2}$ plane. The agents are required to follow a circular trajectory while maintaining a prescribed formation configuration with respect to each other and the virtual leader. The dashed curves represent the agents’ past trajectories, while the solid curves indicate their motion between 50.0 and 54.5 seconds.}
    \label{fig:formation}
\end{figure}

Next, we integrate the estimator (\ref{eq:3}) with the control protocol (\ref{eq:13}) and (\ref{eq:14}) with the parameters chosen as $\gamma_1 = 40$, $\gamma_2 = 1$, $\sigma = 10^{-4}$,  $
H_{i,1} = 900 \times
\bigl[\begin{smallmatrix}
0.8 & 0 & 0 \\
0 & 1 & 0 \\
0 & 0 & 1.5
\end{smallmatrix}\bigr]
$, $
H_{i,2} = 1200 \times 
\bigl[\begin{smallmatrix}
0.8 & 0 & 0 \\
0 & 1 & 0 \\
0 & 0 & 1.5
\end{smallmatrix}\bigr]
    $, and  $ [p_{i1}, p_{i2}, p_{i3},  v_{i1}, v_{i2}, v_{i3}]^T$ as the input of NN for constructing the Gaussian RBF NN $\hat{W}_{k,i}^T S_k(x_i)$ using $4 ^6 = 4069$ neurons with the evenly placed centers over the state space  $[-100, 100] \times [-100, 100] \times [-100, 100] \times[-100, 100]\times [-100, 100] \times [-100, 100]$ and the widths $\gamma_k = 90$  ($\forall i \in \mathbf{I}[1,4]$, $k \in \mathbf{I}[1,3]$).
We simulate the control performance using MATLAB R2024b, with $p_1(0) = [30, 60, 0]^T$,  $p_2(0) = [50, 40, 0]^T$,  $p_3(0) = [50, 80, 0]^T$,  $p_4(0) = [10, 70, 0]^T$ as the agents' initial condition and $\hat{W}_{k,i}(0)=0$ ($\forall i \in \mathbf{I}[1,4]$, $k \in \mathbf{I}[1,3]$). Fig. \ref{fig:positiontracking} illustrate the position tracking performance of the proposed control strategy along the \(p_{i1}\), \(p_{i2}\), and \(p_{i3}\) dimensions with the predefined distance vectors being  $p_1^* = [0, 0, 0]^T$, $p_2^* = [7, -7, 0]^T$, $p_3^* = [14, 0, 0]^T$, and $p_4^* = [7, 7, 0]^T$. Across all three tracking direction agents track the desired trajectory relative to the leader's motion —\(p_{i1} \rightarrow p_{01}\), \(p_{i2} \rightarrow p_{02}\), and \(p_{i3} \rightarrow p_{03}\)— demonstrating highly accurate tracking behavior. The error plots confirm that the tracking errors converge exponentially to a small neighborhood around zero, validating the theoretical guarantees of the control law. The results highlight the robustness and adaptiveness of the controller in achieving coordinated multi-dimensional tracking despite dynamic uncertainties. To visualize how the formation configuration is maintained during trajectory tracking, the agents' navigation paths are plotted in the two-dimensional \(p_{i1}\)–\(p_{i2}\) plane in Fig.\ref{fig:formation}. As evident from the figure, the agents maintain the predefined geometric configuration while collectively converging to and accurately following the desired trajectory over time. Thus, the proposed controller ensures fast convergence of the position tracking errors to a small neighborhood around zero and effectively maintains the desired configuration despite the presence of nonlinearities and system uncertainties.

%%%%%%%%%%%%%%%%%%%%%%%%%%%%%%%%%%%%%%%%%%
\section{Conclusions}

In this study we propose a novel cooperative adaptive formation control scheme using RBF NNs for formation control of a group of nonlinear mechanical systems under complete uncertainty. We generate a diverse reference trajectory considering a virtual leader with time-varying external inputs. We use a cooperative discontinuous nonlinear estimation protocol to estimate the leader's states.  To demonstrate the effectiveness of the proposed control protocol, we first theoretically prove the uniform ultimate boundedness of the overall system and the exponential convergence of the tracking errors to a small neighborhood of zero. We then numerically simulate the formation control performance of the proposed scheme for a group of four MASs. The simulation results demonstrate that the proposed adaptive formation control scheme achieves accurate position tracking while maintaining a predefined geometric configuration, even in the presence of complete system uncertainties.  This characteristic makes the method highly suitable for applications where multiple agents must operate cohesively despite unknown disturbances or dynamic environments. This control approach is particularly relevant to drone formations, where robust coordination under uncertainty is critical—for example, in area coverage, search-and-rescue missions, and cooperative payload transport \cite{c025}. The decentralized nature of the proposed protocol also aligns well with the practical limitations of UAV networks, including limited sensing, communication delays, and real-time computational constraints \cite{c026}. Future work will focus on developing an experience-based, position-swappable formation control protocol.

% \section*{Acknowledgments}

\bibliography{sample}

\end{document}